\documentclass{article}
\usepackage{spconf,amsmath,amssymb,graphicx,hyperref,pifont}
\usepackage{booktabs}
\usepackage{xcolor}

\newcommand{\cmark}{\ding{51}}%
\newcommand{\xmark}{\ding{55}}%

\title{PhonoBridge: Error-Aware Phoneme-Mediated EMG-to-Text Reconstruction}
\name{Wen Qiu$^{1,\ast}$, Jiacheng Zhang$^{1,2,\ast}$, Xuanda Chen$^{1}$,
Siqi Cai$^{2,3,\dagger}$, and Haizhou Li$^{1,2,\dagger}$
\thanks{$^\ast$These authors contributed equally to this work. }
\thanks{$^\dagger$Corresponding authors: Siqi Cai, and Haizhou Li.}
\thanks{This work was supported by the Program for Guangdong Introducing Innovative and Entrepreneurial Teams, Grant No.\ 2023ZT10X044, and by Shenzhen Loop Area Institute under Grant No.\ FPF10120250012. }}
\address{$^{1}$School of Artificial Intelligence, The Chinese University of Hong Kong, Shenzhen, China\\
$^{2}$Shenzhen Loop Area Institute, Shenzhen, China\\
$^{3}$Harbin Institute of Technology, Shenzhen, China}

\begin{document}

\maketitle
\begin{abstract}


Electromyography (EMG) provides a non-invasive means of decoding silently articulated speech into text, but accurate decoding remains challenging due to limited availability of paired EMG–text data. We introduce PhonoBridge, a two-stage framework comprising a channel-aware EMG-to-phoneme frontend and a large language model (LLM) backend for text reconstruction. The frontend is trained on paired EMG–text data that consists of channel-wise feature extraction, cross-channel fusion, and electrode embeddings. With phonemes as the intermediate representation, we expand the backend’s training data by converting LibriSpeech transcripts into phoneme–text pairs without additional EMG recordings. To adapt the LLM to frontend noise, we introduce Error-Profile-Guided Augmentation (EPGA), which perturbs transcript-derived phoneme sequences according to error patterns estimated from frontend predictions, generating additional training pairs for the LLM. On the Open Vocabulary silent-speech dataset, PhonoBridge achieves 24.77\% WER, outperformingthe previously reported best of 28.27\% on the same test split.
\end{abstract}
\begin{keywords}
Silent speech recognition, surface electromyography
\end{keywords}

\section{Introduction}
\label{sec:intro}

Silent-speech interfaces aim to recover linguistic content without requiring an audible utterance. Electromyography (EMG) offers a non-invasive means of recording muscle activity associated with articulation through electrodes placed on the face and neck \cite{review-of-SSI,sEMG-capture-muscle-activity, denby2010silent,stepp2012semg}. This has motivated research on the decoding task, which seeks to recover the intended text or audio from EMG signals.

\begin{figure}[t]
  \centering
  \includegraphics[width=\columnwidth]{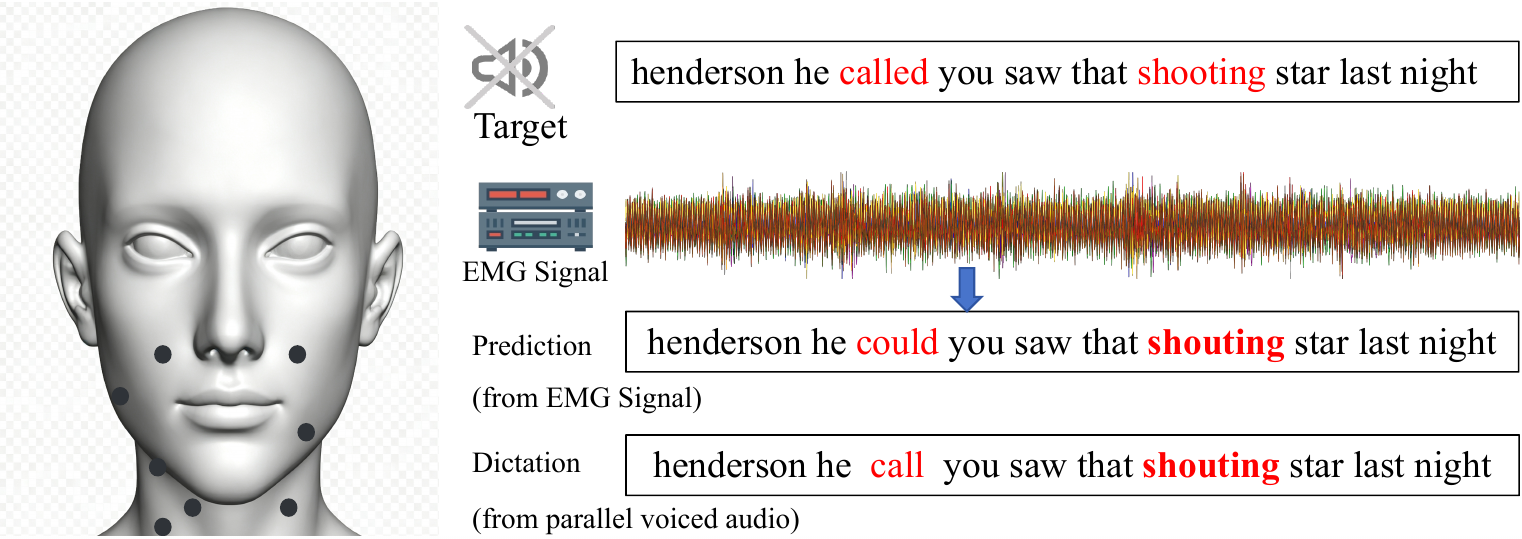}
  \vspace{-5pt}
  \caption{Electrode placement and qualitative comparison of decoding errors. \textbf{Left:} Electrode placement for EMG acquisition. \textbf{Right:} A comparison of EMG-decoded text and human transcription of the corresponding voiced utterance against the reference text.}
  \vspace{-5pt}
  \label{fig:emg-asr}
\end{figure}

Existing studies have adopted modular architectures for EMG-to-text decoding. For example, Gaddy and Klein \cite{gaddy2020digital,gaddy-klein-2021-improved} developed a character-level recognition model trained with connectionist temporal classification (CTC) \cite{ctc}, incorporating an external language model during decoding. Gowda and Miller \cite{gowda-miller-2026-non} used geometric signal representations and CTC training to predict phonemes, followed by weighted finite-state transducer (WFST) decoding with an n-gram language model. 
These methods share a common division of roles: the frontend extracts linguistic information from EMG signals, while the decoder reconstructs the intended text from the frontend’s predictions. Recent advances in large language models (LLMs) offer new opportunities to strengthen the reconstruction stage \cite{qwen3technicalreport} within the same modular framework.

However, an LLM alone does not guarantee accurate reconstruction from EMG-derived predictions. The backend must learn to map noisy frontend outputs to the intended texts, but limited paired EMG-text data provide insufficient supervision for this mapping. External speech transcripts \cite{panayotov2015librispeech} offer abundant training data, but the phoneme sequences derived from them are clean and therefore do not reflect the errors encountered at inference time. This raises a central question: ``\emph{How can transcript-derived phoneme-text pairs be used to train an LLM to correct EMG-specific recognition errors?}''

To address this question, we propose \textbf{PhonoBridge}, a two-stage EMG-to-text framework comprising a channel-aware EMG-to-phoneme frontend and an error-aware LLM-based text reconstruction backend. The frontend learns to predict phoneme sequences from paired EMG–text data using a channel-aware encoder. It first extracts temporal features from each EMG channel separately and then integrates them across channels, with electrode embeddings providing explicit channel identity.

\begin{figure*}[t]
  \centering
  \includegraphics[width=\textwidth]{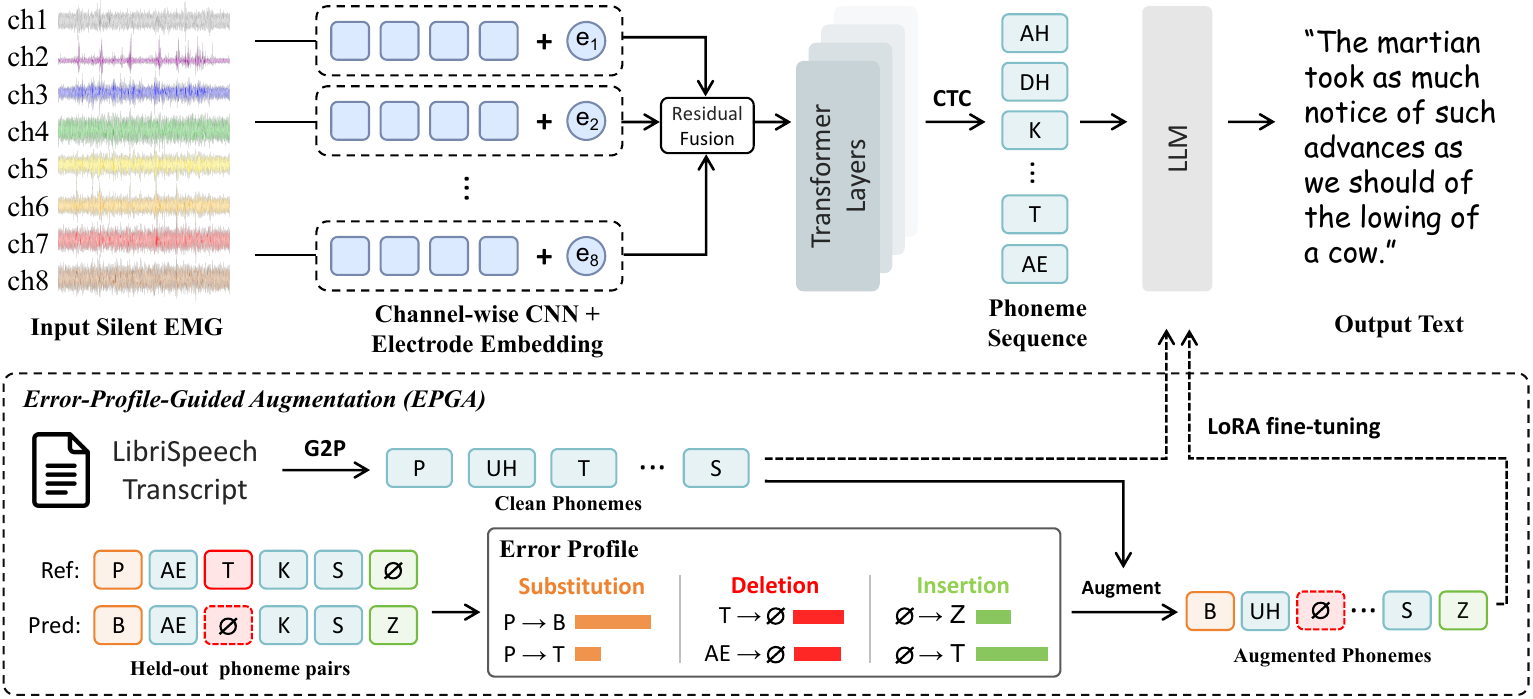}
  \vspace{-5pt}
  \caption{Overall architecture of PhonoBridge.
  The framework comprises two stages: (a) a channel-aware frontend that maps multichannel EMG to phoneme sequences and is trained with CTC loss; and (b) an LLM backend that reconstructs text from the predicted phoneme sequence and is fine-tuned on clean and EPGA-augmented phoneme--text pairs.}
  \vspace{-5pt}
  \label{fig:framework}
\end{figure*}

We use phonemes as both the frontend supervision target and the intermediate representation because EMG and acoustic speech provide different observations of the same underlying articulatory process. During speech production, coordinated muscle activation drives movements of the lips, jaw, and tongue, while EMG records electrical activity associated with these articulatory movements \cite{jou2006continuous}. Although distinct at the signal level, both modalities reflect the same underlying phonemic structure and can therefore exhibit similar phoneme-level confusions. As shown in Fig.~\ref{fig:emg-asr}, both the EMG-decoded output and the human transcription of the parallel voiced utterance replace “shooting” with “shouting.” This shared substitution provides a qualitative example of such cross-modal phonetic similarity and further motivates the use of phonemes as the intermediate representation.

The phoneme interface also allows us to expand the backend’s training data by converting LibriSpeech transcripts into phoneme-text pairs without collecting additional EMG recordings. Since these transcript-derived sequences do not contain the errors produced by the EMG frontend, we introduce Error-Profile-Guided Augmentation (EPGA), which estimates substitution, deletion, and insertion patterns from a held-out subset of the training set and applies them to clean phoneme sequences. We fine-tune the LLM on a mixture of clean and EPGA-augmented phoneme-text pairs, enabling the backend to reconstruct text from noisy EMG-derived predictions. On the Open Vocabulary dataset, PhonoBridge achieves 24.77\% WER, compared with 28.27\% for the strongest prior system on the same test split.

\section{METHOD}
\label{sec:method}

\subsection{Channel-Aware EMG-to-Phoneme Frontend}

Given a multichannel EMG sequence $\mathbf{x}\in\mathbb{R}^{T_0\times n}$ with $T_0$ time samples and $n$ electrode channels, the frontend predicts a phoneme sequence $\hat{\mathbf{y}}$. Since each electrode captures muscle activity from a different facial location, we interleave channel-specific temporal encoding with cross-electrode fusion to preserve electrode-specific information while modeling coordinated muscle activity.

For the $i$-th electrode, let $\mathbf{x}^{(i)}=\mathbf{x}_{:,i}\in\mathbb{R}^{T_0}$. A channel-specific residual convolutional block first extracts and downsamples its local temporal features:
\begin{equation}
    \mathbf{h}^{(i)}=f^{\mathrm{pre}}_i(\mathbf{x}^{(i)})\in\mathbb{R}^{T_1\times d},
\end{equation}
where $T_1$ is the downsampled sequence length, $d$ is the feature dimension, and the branches $f^{\mathrm{pre}}_i$ have independent parameters. Before fusion, each channel is associated with a learnable electrode embedding $\mathbf{e}_i\in\mathbb{R}^{d_e}$. The electrode embedding is concatenated with each channel feature and mapped back to the original feature dimension:
\begin{equation}
    \mathbf{u}^{(i)}_t=\mathbf{W}_e[\mathbf{h}^{(i)}_t;\mathbf{e}_i]+\mathbf{b}_e.
\end{equation}
The resulting channel representations are concatenated across electrodes and passed to a residual temporal fusion module, which jointly models cross-electrode interactions and local temporal context. The fused representation is then sliced into $n$ channel-specific streams, each of which is further processed by two residual convolutional blocks with independent parameters. The refined channel representations are concatenated and projected into the Transformer hidden space.

The resulting sequence is passed through a Transformer encoder \cite{vaswani2017attention,shaw2018relative} with learned relative positional representations to model long-range temporal dependencies.
Let $\mathcal{U}$ denote the target vocabulary.
A linear prediction head produces frame-level logits over the CTC vocabulary $\tilde{\mathcal{U}}=\mathcal{U}\cup\{\langle\mathrm{blank}\rangle\}$, and the frontend is trained with the standard CTC objective \cite{ctc}. During inference, greedy CTC decoding collapses consecutive repetitions and removes the blank symbol, yielding $\hat{\mathbf{y}}$ as input to the LLM backend.

\subsection{LLM-Based Text Reconstruction}

Given the phoneme sequence $\hat{\mathbf{y}}$ decoded by the frontend, we serialize its phoneme units as the LLM input.
The LLM then autoregressively reconstructs the normalized text sequence $\hat{\mathbf{z}}$.

To adapt the backend, we convert transcripts from LibriSpeech~\cite{panayotov2015librispeech} into phoneme sequences $\mathbf{y}$ using grapheme-to-phoneme conversion, producing clean phoneme--text pairs $(\mathbf{y},\mathbf{z})$ without additional EMG recordings. We adapt the LLM with LoRA \cite{hu2021loralowrankadaptationlarge} in two phases: the clean adaptation phase first learns the phoneme-to-text mapping from clean pairs, after which the error-aware adaptation phase continues training on a mixture of EPGA-augmented pairs $(\tilde{\mathbf{y}},\mathbf{z})$ and replayed clean pairs. This exposes the backend to frontend-like errors while retaining its clean reconstruction ability. The construction of $\tilde{\mathbf{y}}$ is described in Sec.~\ref{sec:epga}.

\subsection{Error-Profile-Guided Augmentation}
\label{sec:epga}

EPGA transfers error patterns observed from the frontend to clean transcript-derived phoneme sequences while preserving their text targets.

\noindent\textbf{Error-profile construction.} We freeze the trained frontend, decode a held-out EMG subset using greedy CTC decoding, and align each prediction with its reference by minimum edit distance \cite{levenshtein1966binary}. For each aligned utterance $j$, we record a sentence-level edit-count template $\tau_j=(n_S^{(j)},n_D^{(j)},n_I^{(j)})$, which contains its numbers of substitutions, deletions, and insertions. These templates are grouped into sets $\mathcal{T}_{\ell}$ according to reference length.

We additionally estimate phoneme-conditioned edit statistics. Let $C_{\mathrm{sub}}(a,b)$ count substitutions from reference unit $a$ to predicted unit $b$, $C_{\mathrm{del}}(a)$ count deletions of $a$, $C_{\mathrm{ref}}(a)$ count its occurrences in the reference sequences, and $C_{\mathrm{ins}}(a,b)$ count insertions of $b$ associated with reference context $a$. We define
\begin{equation}
\begin{aligned}
P_{\mathrm{sub}}(b\mid a) &= \frac{C_{\mathrm{sub}}(a,b)}{\sum_{v\neq a}C_{\mathrm{sub}}(a,v)}, &
r_{\mathrm{del}}(a) &= \frac{C_{\mathrm{del}}(a)}{C_{\mathrm{ref}}(a)},\\
P_{\mathrm{ins}}(b\mid a) &= \frac{C_{\mathrm{ins}}(a,b)}{\sum_v C_{\mathrm{ins}}(a,v)}. &&
\end{aligned}
\end{equation}
These statistics characterize replacement content, deletion tendency, and insertion content, respectively.


\noindent\textbf{Profile-guided augmentation.} Given a clean sequence $\mathbf{y}=(y_1,\ldots,y_L)$, EPGA identifies its length bucket $\ell$ and samples an edit-count template $\tau=(n_S,n_D,n_I)$ from $\mathcal{T}_{\ell}$. The template specifies the numbers of substitutions, deletions, and insertions, preserving the sentence-level edit-count structure observed in frontend predictions of similar length.

To instantiate the template, EPGA first samples $n_D$ deletion positions without replacement according to $r_{\mathrm{del}}(y_i)$. It then selects $n_S$ substitution positions from the remaining reference units and draws a replacement $b$ for each selected position from $P_{\mathrm{sub}}(b\mid y_i)$. For each of the $n_I$ insertions, an anchor is sampled from the sentence start and the reference positions. Given its reference context $a$, the inserted unit $b$ is drawn from $P_{\mathrm{ins}}(b\mid a)$. The resulting sequence $\tilde{\mathbf{y}}=\mathcal{A}_{\mathrm{EPGA}}(\mathbf{y})$ is paired with the original text target.

\section{EXPERIMENTS}


\subsection{Experimental Setting}
\textbf{Dataset.}
We evaluate on the eight-channel Open Vocabulary dataset introduced by Gaddy and Klein~\cite{gaddy2020digital}, using the official split and text normalization protocol of~\cite{gaddy-klein-2021-improved}. We reserve a held-out subset of the official training split exclusively for estimating the EPGA error profile and use the remaining data to train the frontend. We report phoneme error rate (PER) for frontend recognition and word and character error rates (WER/CER) for reconstruction and end-to-end evaluation on the test set. Lower error rates are preferred.

\noindent\textbf{Baselines.}
We compare PhonoBridge with Gaddy and Klein~\cite{gaddy-klein-2021-improved},  EMG2Vec~\cite{10781736}, EMG-LLM~\cite{mohapatra2025llms}, and two word-level CTC~\cite{ctc} baselines using CNN-Transformer~\cite{vaswani2017attention} and CNN-BiLSTM~\cite{hochreiter1997long} encoders. All systems are evaluated on the same test split.

\subsection{Implementation Details}

\textbf{EMG-to-phoneme frontend.}
The 1-kHz EMG signals are notch-filtered at 60\,Hz and resampled to 800\,Hz. The frontend is initialized from a character-target warm-start checkpoint and trained with CTC using AdamW \cite{loshchilov2019decoupled} ($3\times10^{-4}$); the checkpoint with the lowest development-set PER is used for subsequent experiments.

\noindent\textbf{LLM backend adaptation.}
We adapt Qwen3-8B \cite{qwen3technicalreport} with LoRA  (rank 16) in two phases. The clean adaptation phase uses 1M LibriSpeech-derived phoneme--text pairs, followed by an error-aware adaptation phase using 1M EPGA-augmented pairs mixed with 200k clean replay pairs. Each phase runs for one epoch with learning rates of $2\times10^{-4}$ and $1\times10^{-4}$, respectively. 
EPGA estimates length-conditioned, phoneme-dependent edit statistics from held-out reference--prediction alignments, with the sampled error rate scaled by 1.17 and capped at 0.30. 
We use greedy decoding for inference.

\section{RESULTS AND DISCUSSION}


\subsection{Main Results}


\begin{table}[t]
\centering
\caption{End-to-end WER (\%) on the Open Vocabulary test split. All systems are evaluated on the same split.} 
\label{tab:main}
\begin{tabular}{lc}
\hline
Method & WER (\%) \\
\hline
Gaddy and Klein~\cite{gaddy-klein-2021-improved} & 28.27 \\
EMG2Vec~\cite{10781736} & 29.45 \\
EMG-LLM~\cite{mohapatra2025llms} & 108.29 \\
CNN-Transformer & 51.29 \\
CNN-BiLSTM & 45.31 \\
\textbf{PhonoBridge (Ours)} & \textbf{24.77} \\
\hline
\end{tabular}
\vspace{-8pt}
\end{table}

Table~\ref{tab:main} compares PhonoBridge against baseline systems. PhonoBridge achieves 24.77\% WER, outperforming all baselines, including the previous state-of-the-art Gaddy and Klein system (28.27\%) and the self-supervised EMG2Vec approach by Hou et al. (29.45\%). The substantial improvement over EMG-LLM by Mohapatra et al. (108.29\% WER) supports the effectiveness of phoneme-mediated decoupling over direct EMG-to-text approaches. Our method also substantially outperforms the direct CNN-based architectures (CNN-Transformer and CNN-BiLSTM), providing further evidence for the two-stage design that combines signal-level encoding with linguistic reconstruction.

\subsection{Channel-Aware Encoder Ablation}

We ablate the channel-aware design in the frontend encoder to validate the effectiveness of channel-wise processing and electrode embeddings. Table~\ref{tab:channel} compares different frontend configurations on PER and collapse WER (cWER), computed by splitting predicted and reference phoneme sequences at word-boundary tokens (\texttt{<SEP>}) and measuring edit distance \cite{levenshtein1966binary} between the resulting phoneme chunks.

\begin{table}[t]
\centering
\caption{Channel-aware encoder ablation on the test split. Results are reported in PER and collapse WER (cWER), with Gaddy and Klein included for reference.}
\label{tab:channel}
\footnotesize
\setlength{\tabcolsep}{3pt}
\begin{tabular}{lcccc}
\toprule
Method & Channel-wise & Elec.\ emb. & PER (\%) & cWER (\%) \\
\midrule
Gaddy and Klein~\cite{gaddy-klein-2021-improved}
    & \xmark & \xmark & 14.12 & 36.35 \\
\midrule
PhonoBridge
    & \xmark & \xmark & 33.43 & 67.35 \\
PhonoBridge
    & \xmark & \cmark & 32.39 & 63.76 \\
PhonoBridge
    & \cmark & \xmark & 14.17 & 37.62 \\
\textbf{PhonoBridge}
    & \cmark & \cmark & \textbf{13.65} & \textbf{35.53} \\
\bottomrule
\end{tabular}
\vspace{-8pt}
\end{table}

Channel-wise processing provides the largest improvement: without electrode embeddings, it reduces PER from 33.43\% to 14.17\% and cWER from 67.35\% to 37.62\%. Electrode embeddings provide additional gains, with the complete encoder achieving the lowest PER (13.65\%) and cWER (35.53\%), slightly outperforming Gaddy and Klein on both metrics. These results indicate that preserving channel-specific patterns before cross-channel fusion is critical, while explicitly encoding electrode identity provides a further improvement.

\subsection{Backend Training Strategy Ablation}
We ablate the backend adaptation strategy to examine the roles of clean and EPGA-augmented phoneme--text pairs and the benefit of sequential training. Table~\ref{tab:backend_ablation} compares zero-shot inference, single-source adaptation, single-phase mixed adaptation, and our two-phase procedure.

\begin{table}[t]
\centering
\caption{Ablation of LLM backend adaptation on the test set. Mixed training combines EPGA-augmented pairs with replayed clean pairs, and $\rightarrow$ denotes sequential adaptation.}
\label{tab:backend_ablation}
\small
\begin{tabular}{lcc}
\toprule
Adaptation setting & WER (\%) & CER (\%) \\
\midrule
Qwen3-8B (zero-shot)       & 188.22 & 116.77 \\
\midrule
Clean only                 & 66.53  & 34.53  \\
EPGA-augmented only        & 29.41  & 17.23  \\
Mixed, single phase        & 25.89  & 22.39  \\
\textbf{Clean $\rightarrow$ mixed (ours)}
                           & \textbf{24.77} & \textbf{14.58} \\
\bottomrule
\end{tabular}
\vspace{-8pt}
\end{table}

Without task-specific adaptation, Qwen3-8B performs poorly on phoneme-to-text reconstruction. Clean-only adaptation learns the basic phoneme-to-text mapping but remains sensitive to frontend errors, whereas training only on EPGA-augmented pairs improves robustness but is less effective than combining the two data sources. Direct single-phase training on the mixed data achieves 25.89\% WER, while our two-phase strategy further reduces WER to 24.77\% and CER to 14.58\%. This improvement indicates that both the training-data composition and adaptation order contribute to the final performance.





\section{CONCLUSION}
This paper presents PhonoBridge, a phoneme-mediated framework for open-vocabulary silent-speech decoding from multichannel EMG. It separates phoneme recognition from text reconstruction and uses EPGA to expose the language model to realistic frontend errors. Experiments on the Open Vocabulary dataset demonstrated its effectiveness, with ablations confirming the contributions of channel-aware processing, electrode embeddings, EPGA, and sequential backend adaptation. Overall, the results support phoneme-level mediation for connecting noisy EMG recognition with open-vocabulary text generation.



\bibliographystyle{IEEEbib}
\newpage
\bibliography{strings,refs}

\begin{thebibliography}{10}

\bibitem{review-of-SSI}
Jose~A. Gonzalez-Lopez, Alejandro Gomez-Alanis, Juan~M. Martín~Doñas,
  José~L. Pérez-Córdoba, and Angel~M. Gomez,
\newblock ``Silent speech interfaces for speech restoration: A review,''
\newblock {\em IEEE Access}, vol. 8, pp. 177995--178021, 2020.

\bibitem{sEMG-capture-muscle-activity}
Xinyi Yang, Jing Yan, Hanting Zhang, and Liang Wu,
\newblock ``Noninvasive and quantitative measurement of intrinsic laryngeal
  muscle activity using surface electromyography,''
\newblock {\em IEEE Transactions on Instrumentation and Measurement}, vol. 74,
  pp. 1--10, 2025.

\bibitem{denby2010silent}
Bradley Denby, Tanja Schultz, Kiyoshi Honda, Thomas Hueber, Juan~M. Gilbert,
  and Jonathan~S. Brumberg,
\newblock ``Silent speech interfaces,''
\newblock {\em Speech Communication}, vol. 52, no. 4, pp. 270--287, 2010.

\bibitem{stepp2012semg}
Cara~E. Stepp,
\newblock ``Surface electromyography for speech and swallowing systems:
  Measurement, analysis, and interpretation,''
\newblock {\em Journal of Speech, Language, and Hearing Research}, vol. 55, no.
  4, pp. 1232--1246, 2012.

\bibitem{gaddy2020digital}
David Gaddy and Dan Klein,
\newblock ``Digital voicing of silent speech,''
\newblock in {\em Proceedings of the 2020 Conference on Empirical Methods in
  Natural Language Processing}. 2020, pp. 5521--5530, Association for
  Computational Linguistics.

\bibitem{gaddy-klein-2021-improved}
David Gaddy and Dan Klein,
\newblock ``An improved model for voicing silent speech,''
\newblock in {\em Proceedings of the 59th Annual Meeting of the Association for
  Computational Linguistics and the 11th International Joint Conference on
  Natural Language Processing (Volume 2: Short Papers)}. 2021, pp. 175--181,
  Association for Computational Linguistics.

\bibitem{ctc}
Alex Graves, Santiago Fern{\'a}ndez, Faustino Gomez, and J{\"u}rgen
  Schmidhuber,
\newblock ``Connectionist temporal classification: labelling unsegmented
  sequence data with recurrent neural networks,''
\newblock in {\em Proceedings of the 23rd International Conference on Machine
  Learning}, New York, NY, USA, 2006, ICML '06, pp. 369--376, Association for
  Computing Machinery.

\bibitem{gowda-miller-2026-non}
Harshavardhana~T Gowda and Lee~M. Miller,
\newblock ``Non-invasive electromyographic speech neuroprosthesis: a geometric
  perspective,''
\newblock in {\em Findings of the {A}ssociation for {C}omputational
  {L}inguistics: {ACL} 2026}, Maria Liakata, Viviane~P. Moreira, Jiajun Zhang,
  and David Jurgens, Eds., San Diego, California, United States, July 2026, pp.
  11636--11650, Association for Computational Linguistics.

\bibitem{qwen3technicalreport}
An~Yang, Anfeng Li, Baosong Yang, et~al.,
\newblock ``Qwen3 technical report,'' 2025.

\bibitem{panayotov2015librispeech}
Vassil Panayotov, Guoguo Chen, Daniel Povey, and Sanjeev Khudanpur,
\newblock ``Librispeech: An asr corpus based on public domain audio books,''
\newblock in {\em 2015 IEEE International Conference on Acoustics, Speech and
  Signal Processing}, 2015, pp. 5206--5210.

\bibitem{jou2006continuous}
Szu-Chen Jou, Tanja Schultz, Matthias Walliczek, Florian Kraft, and Alex
  Waibel,
\newblock ``Towards continuous speech recognition using surface
  electromyography,''
\newblock in {\em Proceedings of Interspeech 2006}, 2006, pp. paper
  1592--Mon3WeS.3.

\bibitem{vaswani2017attention}
Ashish Vaswani, Noam Shazeer, Niki Parmar, Jakob Uszkoreit, Llion Jones,
  Aidan~N. Gomez, Lukasz Kaiser, and Illia Polosukhin,
\newblock ``Attention is all you need,''
\newblock in {\em Advances in Neural Information Processing Systems 30}, 2017,
  pp. 5998--6008.

\bibitem{shaw2018relative}
Peter Shaw, Jakob Uszkoreit, and Ashish Vaswani,
\newblock ``Self-attention with relative position representations,''
\newblock in {\em Proceedings of the 2018 Conference of the North American
  Chapter of the Association for Computational Linguistics: Human Language
  Technologies, Volume 2 (Short Papers)}. 2018, pp. 464--468, Association for
  Computational Linguistics.

\bibitem{hu2021loralowrankadaptationlarge}
Edward~J. Hu, Yelong Shen, Phillip Wallis, Zeyuan Allen-Zhu, Yuanzhi Li, Shean
  Wang, Lu~Wang, and Weizhu Chen,
\newblock ``Lora: Low-rank adaptation of large language models,'' 2021.

\bibitem{levenshtein1966binary}
Vladimir~I. Levenshtein,
\newblock ``Binary codes capable of correcting deletions, insertions, and
  reversals,''
\newblock {\em Soviet Physics Doklady}, vol. 10, pp. 707--710, 1966.

\bibitem{10781736}
Qinhan Hou, Stefano van Gogh, Kevin Scheck, Zhao Ren, Tanja Schultz, Michael
  Wand, and Jürgen Schmidhuber,
\newblock ``emg2vec: Self-supervised pretraining in electromyography-based
  silent speech interfaces,''
\newblock in {\em 2024 46th Annual International Conference of the IEEE
  Engineering in Medicine and Biology Society (EMBC)}, 2024, pp. 1--5.

\bibitem{mohapatra2025llms}
Payal Mohapatra, Akash Pandey, Xiaoyuan Zhang, and Qi~Zhu,
\newblock ``Can {LLM}s understand unvoiced speech? exploring {EMG}-to-text
  conversion with {LLM}s,''
\newblock in {\em Proceedings of the 63rd Annual Meeting of the Association for
  Computational Linguistics}. 2025, pp. 703--712, Association for Computational
  Linguistics.

\bibitem{hochreiter1997long}
Sepp Hochreiter and J{\"u}rgen Schmidhuber,
\newblock ``Long short-term memory,''
\newblock {\em Neural Computation}, vol. 9, no. 8, pp. 1735--1780, 1997.

\bibitem{loshchilov2019decoupled}
Ilya Loshchilov and Frank Hutter,
\newblock ``Decoupled weight decay regularization,''
\newblock in {\em International Conference on Learning Representations}, 2019.

\end{thebibliography}

\end{document}